\documentclass{scrartcl}
\usepackage[latin9]{inputenc}
\usepackage[T1]{fontenc}
\usepackage{placeins}

\usepackage{amsmath, amssymb, amsfonts}
\usepackage{dsfont}
\usepackage[all]{xy}
\usepackage{amsthm}
\usepackage{enumerate}
\usepackage{tikz}
\usepackage{todonotes}
\usepackage{stmaryrd}
\usetikzlibrary{matrix}
\usetikzlibrary{trees}
\usepackage[hyphens]{url}
\usepackage{caption}
\usepackage{graphicx}
\usepackage{natbib}
\usepackage{todonotes}
\usepackage{chngcntr}
\counterwithin{figure}{section}
\counterwithin{table}{section}
\usepackage{float}
\usepackage{enumitem,xcolor}

\makeatletter
\renewcommand*\env@matrix[1][*\c@MaxMatrixCols c]{%
  \hskip -\arraycolsep
  \let\@ifnextchar\new@ifnextchar
  \array{#1}}
\makeatother

\theoremstyle{plain}

\theoremstyle{definition}

\theoremstyle{remark}

\usepackage{blindtext} 
\usepackage{framed} 
\usepackage{xcolor} 
\colorlet{shadecolor}{gray!25} 
	
\begin{document}

\title{A copula-based time series model for global horizontal irradiation}
\author{ Alfred M\"{u}ller and Matthias Reuber
\\
Department of Mathematics,\\
University of Siegen, Germany. \\
\texttt{mueller,reuber@mathematik.uni-siegen.de}\\}
\date{\today }

%
%
\maketitle
%

\section*{Abstract}
The increasing importance of solar power for electricity generation leads to an increasing demand for probabilistic forecasting of local and aggregated PV yields. In this paper we use an indirect modeling approach for hourly medium to long term local PV yields based on publicly available irradiation data. We suggest a time series model 
for global horizontal irradiation for which it is easy to generate an arbitrary number of scenarios and thus allows for multivariate probabilistic forecasts for arbitrary time horizons. In contrast to many simplified models that have been considered in the literature so far it features several important stylized facts. Sharp time dependent
 lower and upper bounds of global horizontal irradiations are estimated that improve the often used physical bounds. The parameters of the beta distributed marginals of the transformed data are allowed to be time dependent. A copula-based time series model is introduced for the hourly and daily dependence structure based on a simple graphical structure known from the theory of vine copulas. Non-Gaussian copulas like Gumbel and BB1 copulas are used that allow for the important feature of so-called tail dependence. Evaluation methods like the continuous ranked probability score (CRPS), the energy score (ES) and the variogram score (VS) are used to compare the power of the model for multivariate probabilistic forecasting with other models used in the literature showing that our model outperforms other models in many respects. 
\section*{Keywords}
probabilistic solar power forecasting, global horizontal irradiation, probabilistic scenario generation, copula models, time series model, tail dependence, scoring rules

\section{Introduction}
Over the past decades the global importance of electricity provided by renewable energies has been continuously increasing. 
Worldwide installed capacities of photovoltaics have increased from 9 GW in 2007 to more than 500 GW in 2018 according to the \textit{International Renewable Energy Agency}.\\
Photovoltaic yields depend on the irradiation and further weather influences and are only positive between sunrise and sunset. 
Much of the variability is predictable given the position of the sun. With formulas for the rotational and translational movements of the earth with respect to the sun we can explain these predictable parts of the fluctuations. Unexpected changes occur mainly because of the presence of clouds. 

In this paper we will develop a stochastic model that will enable us to generate scenarios of global horizontal irradiation and thus also of photovoltaic yields over a medium-term horizon of up to a few years. 
Such models have many possible applications dependent on the level they are used. Models for local plants can help to evaluate the financial benefit of a PV plant. Combined with load models the stochastic grid infeeds and the own consumption can be calculated. This supports investors to estimate the value of a PV plant or a purchasing power agreement (PPA) and to choose the optimal sizing of a battery storage. Considering global PV yields (e.g. the area of a transmission system operator) such probabilistic models can support the grid size planning and moreover, help to analyze the future financial value of PV energy from a societal point of view and its effect on electricity prices. We mention a few recent studies, where such problems are adressed and where such a model could be helpful. 

In this paper we propose a probabilistic model for hourly global horizontal irradiation that can also be applied to local or global PV yields. First, a statistical estimation method of lower and upper bounds is introduced. The necessity of an estimated upper bound has recently also been observed by 
\cite{woodruff2018} in the context of short-term probabilistic forecasts of solar power generation. We are not aware of any investigation of this problem in a time
series context, however. 

Assuming the knowledge of sharp upper and lower bounds we then consider copula-based time series models to describe the hourly and daily dependence structure. It is natural to assume a time dependent beta distribution for the univariate distributions. This is a very common approach used in the literature, see e.g. \cite{hasan2019} for an overview of the literature with many references. 
Copulas have already been used to describe the dependence structure of PV yields and wind yields in several other studies. Very often a Gaussian copula is assumed.  \cite{golastaneh2016} describe the space-time dependencies of hourly PV yields with multivariate Gaussian copulas. Gaussian copulas are also used in \cite{pinson2012} for short-term wind models and in \cite{pinson2013} for space-time dependencies of wind yields.
\cite{golastaneh2017} use R-vine and Gaussian copulas to generate multivariate prediction intervals.  

None of these papers deals with the concept of so-called \textit{tail dependence} as we do in this paper. Our data show that tail dependence should be taken into account and that it is quite significant in particular around noon. Unfortunately, a well known mathematical result shows that a time series model based on a Gaussian copula can never exhibit tail dependence. Therefore other copula models have to be considered to take that feature into account. We therefore suggest here to consider Gumbel and BB1 copulas to describe dependence structures allowing for tail dependence. 

This paper of course is not the first one to consider time series models for longer term PV yields. Models of this kind are already used in \cite{wagner2014}, \cite{veraart2015}, \cite{ibrahim2017} and \cite{benth2017}, to mention a few. All of them put a special focus on daily mean or maximum PV yields and only use very simple deterministic approaches to obtain hourly PV yields. The intraday behavior is considerable better described by a copula approach that we introduce in this paper. We use the structure of a Markov tree which is a simple version of a so-called vine copula. The idea of using vine copulas for PV yields
was previously used in \cite{golastaneh2017}  in the context of short-term probabilistic forecasting and by \cite{wu2015} where a vine copula model was suggested for PV generation at different locations at a fixed point of time. Using vine copulas in a time series context for modelling longer term PV yields seems to be new. A copula-based Bayesian method for probabilistic solar power forecasting is introduced in \cite{panamtasch2020}.

To close this literature review section we want to mention a few recent review articles on irradiation and PV models. A general overview of the existing literature in the area of probabilistic forecasting of solar power is given by \cite{vandermeer2018b}. Caused by the Global energy forecasting competition (GEFCom) 2014 (see \cite{hong2016}) there has been an increase in the interest of probabilistic models of solar energy. The contestants of the solar track were supposed to submit forecasts in the form of 99 quantiles within a rolling forecasting of three solar farms at different locations. Out of the competition a huge number of papers arose. For example, \cite{nagy2016} and \cite{juban2016} deploy a quantile regression in combination with different approaches based on generalized additive trees and radial basis functions. We refer to \cite{hong2014} and \cite{hong2019} for further information on the GEFCom 2012 and 2017 that also include energy and probabilistic forecasting. 

The evaluation of forecasts in these studies typically concentrates on evaluating a number of quantiles of the distributions at fixed times and thus evaluating only the marginal distributions and thus do not assess the quality of the models for forecasting multivariate distributions, which is also of importance in many applications. In their seminal paper on scoring rules for probabilistic forecasting, \cite{gneiting2012} suggest the \textit{energy score} as a strictly proper scoring rule for multivariate distributions that we also consider in this paper. It is known that it is difficult to assess dependence properties, in particular tail dependence, as has been discussed recently e.g. in \cite{brehmer2019}. Therefore we will also consider other approaches to evaluate our multivariate forecasts like considering special tail events that one wants to forecast.

The rest of this paper is organized as follows. In section 2 we look at the data used in this article and describe the stylized facts of hourly global horizontal irradiation. The methodology of our model is introduced in section 3. We address the estimation of upper and lower bounds, the choice of hourly marginal distributions and the copula as well as scoring rules for evaluating probabilistic forecasts. Empirical results are shown in section 4. Section 5 concludes.

\section{Irradiation data and its stylized facts}
The irradiation data used in this article is provided by the Copernicus Atmosphere Monitoring Service \cite{cams2018}. 
It provides miscellaneous irradiation values on the ground level. Time series data of different types of irradiation both for actual and clear sky conditions are available. 
As a concrete example we use in this paper in particular hourly global horizontal irradiation (GHI) and the irradiation at the top of the atmosphere (TOA) from 2005 to 2018 for our hometown Siegen in Germany. 
TOA is a function of the latitude, the number of the day in the year and the time during the day. It can be calculated with the solar constant and physical formulas as described in \cite{duffie2006}. An overview of the available data and further information is given in \cite{schroedter2017}. We also want to mention that GHI can be decomposed into beam and diffuse irradiation. While beam irradiation comes directly from the sun to the ground of the earth, diffuse irradiation is reflected either by clouds, the ground or the surroundings. To compute the PV yield for a single plant one needs the distinction into both irradiation types. In this article we only focus on stochastic models for GHI. However, the models introduced here can easily be adopted to local, regional or global PV yields with only small adjustments.\\

\begin{figure}[h]
	\begin{center}
		\includegraphics[width=\textwidth]{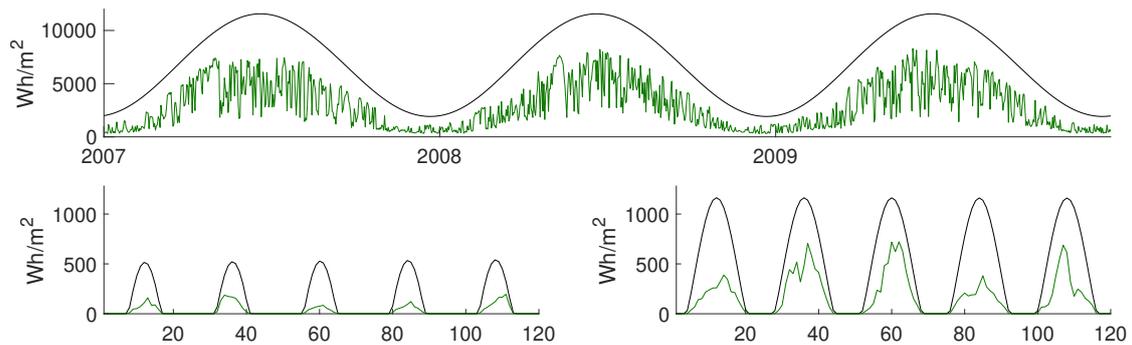}
	\end{center}
	\caption{Three years of daily GHI and TOA (above) and five days of hourly GHI and TOA in winter (bottom left) respectively summer (bottom right).}
	\label{irradiation}
\end{figure}

A graphical representation of the time series of GHI on daily and hourly levels is presented in figure 2.1. Three years of daily GHI and TOA are shown in the upper picture and five days of hourly GHI and TOA are shown in the lower pictures for days in December and June. In the following we present the stylized facts of hourly GHI. All values of hourly GHI are \textbf{positive} in the course of a solar day, that is the time between sunrise and sunset. The day length is changing in the course of the year with longer days in summer than in winter on the northern hemisphere. \\
There is a \textbf{yearly seasonality}. Hourly values of GHI for a given hour take in general higher values in the summer and lower values in the winter. This is due to the changing position of the sun in the course of the year and the thereof resulting different day lengths. Besides the yearly seasonality there is also a \textbf{daily seasonality}. The changing sun height affects the amount of global irradiation reaching the Earth's surface in the course of the day. Higher values are observed at noon, when the sun reaches its highest position. In fact, both the yearly and the daily seasonality of GHI are compositions of two types of seasonalities. On the one hand there is the seasonality that arises through the changing position and angle of the sun. This is expressed by the changing day length and the higher average irradiation values in summer and around noon. An additional seasonality is due to changes in the cloud behavior. In summer there are on average less clouds than in winter and so the irradiation is more often close to its upper bound than in winter. We will address the first seasonality by upper and lower bounds and the second seasonality by a time dependent beta distribution for hourly marginals. \\
TOA is a natural upper bound and zero a natural lower bounds for GHI. However, the values of GHI are limited by a time-varying \textbf{upper bound} which seems to be lower than the TOA. There is a considerable deviation between GHI and TOA as a part of the irradiation is always absorbed on its way through the atmosphere. Aerosols, water vapor and turbidity lead to the fact, that not the whole irradiation reaches the Earth's surface. Moreover, also in the case of a total overcast sky there is a positive irradiation, so that there seems to be a seasonal \textbf{lower bound} strictly greater than zero. The lower bound results from the diffuse irradiation that is always greater than zero (we exclude volcanic eruptions and solar eclipses) between sunrise and sunset, even in the case of a total overcast sky. Estimation of these bounds is explained in detail in section 3.1. \\
There is also an \textbf{intraday variability} with a magnitude depending on the total daily irradiation. This variability is due to the cloud movement. More frequent changes in the weather situations lead to a higher variability. In the case of a high total daily irradiation the intraday variability is much lower than in the case of a medium total daily irradiation. As a last characteristic of GHI there is a \textbf{time dependence} on daily and hourly levels that is due to relatively stable weather conditions in the near term. As for the variability this dependence is stronger in case of a high daily irradiation compared to the case of a medium value for the total daily irradiation.\\

\begin{figure}[h]
	\begin{center}
		\includegraphics[width=\textwidth]{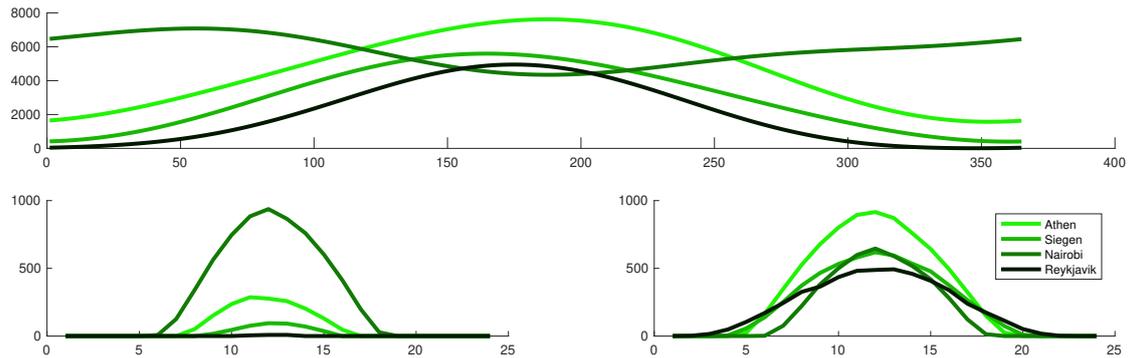}
	\end{center}
	\caption{Daily seasonality function (above) and hourly seasonality functions for winter (bottom left) respectively summer (bottom right).}
	\label{irradiation_all}
\end{figure}

In this paper we mainly consider the location of Siegen, but we want to point out that other locations exhibit the same stylized facts. They only differ in some aspects due to different angles of the sun and climate conditions what leads to different parameter values. We demonstrate this by considering the yearly and daily seasonalities of four different locations that are Athens, Nairobi, Reykjavik and Siegen. The expected irradiation for every hour $h$ can be modeled with a truncated Fourier series approach depicting yearly and half-yearly seasonalities
\begin{align*}
			d \mapsto \mathbb{E}(G_{d,h})=\beta_{0,h}+\sum_{i=1}^2\beta_{i,h}\cos\left(\frac{2\pi d i}{365.25}\right)+\sum_{j=3}^4\beta_{j,h}\sin\left(\frac{2\pi d j}{365.25}\right).
\end{align*}
In section 3.1 we will consider such a function in the context of a quantile regression to estimate lower and upper bounds of hourly GHI. The daily and hourly seasonality functions are shown in figure \ref{irradiation_all}. In the course of the year we can see differences in the general level of the irradiation and the time of the year with maximum irradiation values. Changing day lengths as well as different levels of GHI can also be seen in the bottom two plots of figure \ref{irradiation_all}. The different behavior of the seasonality functions of the four locations is also reflected in the parameters of the linear regression. Table \ref{t:parameters} shows an example of the different parameters for the hour $h=12$.

\begin{table}[h]
\centering
	\captionabove[Parameters]{Overview of the seasonality parameters ($h=12$) for the four different locations.}
	\label{t:parameters}
	\begin{tabular}{|	l	||	lllll	|}
	\hline
		&$\beta_{0,12}$&$\beta_{1,12}$&$\beta_{2,12}$&$\beta_{3,12}$&$\beta_{4,12}$\\
	\hline
	\hline
		Athens&616.3&-321.42&-20.03&19.03&32.91\\
		Nairobi&848.6&145.15&-58.28&63.53&24.72\\
		Reykjavik&234.08&-239.63&13.04&38.91&2.08\\
		Siegen&383.41&-260.69&-28.78&55.83&-9.7\\
	\hline
	\end{tabular}
\end{table}  
  
In the following our main focus is on the GHI of Siegen. However, the models that will be introduces can be applied to different further locations, which only changes the values of the parameters of the model. In the rest of the paper we will try to find models that exhibit all the discussed stylized facts.

\section{Methodology}
Let $G_t$ be the global horizontal irradiation at a certain time $t$. The fundamental model equation for GHI is given as:
\begin{align*}
	G_t&=g_t^-+M_t\cdot (g_t^+-g_t^-),
\end{align*}
where
\begin{itemize}
	\item $g_t^-$ and $g_t^+$ are the lower and upper bounds of global horizontal irradiation and
	\item $M_t\in (0,1)$  can be considered as the intensity of the sun.
\end{itemize}
An alternative equivalent formulation of the model equation is given by
\begin{align*}
	G_t&=g_t^+-(1-M_t)\cdot (g_t^+-g_t^-).
\end{align*}
In this case $(1-M_t)$ can be interpreted as the cloud coverage. In both formulations the cloud behavior that influences the global horizontal irradiation is modeled by $M_t$. 

In the following subsections we will describe the details of this model. First we will determine estimates of the upper and lower bounds $g_t^-$ and $g_t^+$ in subsection 3.1. Then we will consider appropriate marginal distributions for $M_t$ in subsection 3.2. The dependence structure of the $M_t$ for different $t$ will be described with the help of copulas in subsection 3.3 und finally we will describe in subsection 3.4 how
one can compare the quality of different model approaches with evaluation methods based on scoring rules.

From now on we will very often write $t = (d,h)$ for the time index, where $d$ will denote the day and $h$ the hour of the day. 
We will also use vector notation and will write $\mathbf{G}_d=(G_{d,0},\dotsc,G_{d,23})^T$ for the vector of hourly irradiations $G_{d,h}$ at day $d$.

\subsection{Estimation of upper and lower bounds}
On its way through the atmosphere irradation is reduced through absorptions and scatterings. So it is hardly surprising that the quantity TOA of the irradiation at the top of the atmosphere  
is only a very rough upper bound for GHI that is far away from the possible values of GHI as one can see in figure 2.1. As an alternative to TOA to get a better upper bound one could think
of so-called  \textit{clear sky models} that are used in the literature. Clear sky models consider the irradiation reaching the Earths surface in the case of no clouds and therefore are close to observed maximum values. An overview and comparison of six different clear sky models is given in \cite{ineichen2006}. Just to name a few examples there are the Solis and the Kasten model. The first one needs the ozone depth, the water vapor and the aerosol optical depth as input parameters while the latter one needs the Linke turbidity at air mass 2 as an input by taking the absorption and diffusion at different altitudes into account. \cite{bacher2009} present a method based on a quantile regression with Gaussian kernels as a method for computing the clear sky index without further input parameters, but their clear sky irradiation can be exceeded by the true irradiations. 

There are several shortcomings concerning clear sky models. They need further exogenous input variables, that are not always available for every location in a good quality. Moreover, the clear sky models are difficult to generalize for regional (e.g. the area of a transmission system operator) or global (e.g. whole Germany) PV yields. We want to introduce a model that can be used both for irradiations and PV yields. Most importantly, however, clear sky irradiations can be exceeded by observed GHI values. A physical justification for this phenomenon lies in different levels of water vapor and the reflection of irradiation through thin clouds that can enhance the irradiation in particular around sunrise and sunset.

Therefore we propose a statistical estimation method for an upper and a lower bound for hourly GHI that is based only on the observed GHI values and has the characteristic that all observations lie between the lower and upper bound. There is not much literature on statistical estimation methods for such bounds. \cite{daouia2017} present an R package for nonparametric boundary regression and provide an overview of existing approaches. Their goal is rather to estimate the boundary of a bivariate distribution than to estimate the boundaries of a time series. We use an approach based on a quantile regression and the peak over threshold (POT) method. Quantile regression (\cite{koenker1978}) is a well known method to estimate quantiles. We need the limit of $\tau$-quantiles when the level $\tau$ approaches $\tau \to 1$. 
The idea is to model the shape of the upper bound with a quantile regression for a high quantile but to shift it upwards in a meaningful way. The shifting upwards will be done with
a well known method from extreme value statistics to estimate the right endpoint of a distribution. We will use the so-called peak over threshold (POT) method to model extreme events that exceed a given high threshold with a generalized Pareto distribution (see \cite{coles2001}). We first describe the procedure in the case of an upper bound. Exceedances of the quantiles are then used to estimate the right endpoint of a generalized Pareto distribution (GPD). An upper bound is obtained by pushing the estimation of the quantile upwards with the estimated right endpoint of the GPD. 
\\
We now describe the steps of the algorithm for the upper bound. First, for each fixed day $d$ and fixed hour $h$ of the year we compute the historical maximum irradiation values 
\begin{align*}
	\tilde{g}_{d,h}^+=\max_{i=1,\dotsc,n}g_{d,h}^{(i)},\quad \ d=1,\dotsc,365, \ h=0,\dotsc,23,
\end{align*}
for the observed $n$ years in the learn period. Here $g_{d,h}^{(i)}$ is the GHI in year $i$ on day $d$ and hour $h$. Then for all fixed $h\in\{0,\dotsc,23\}$ with positive expected irradiations and a fixed quantile level $\tau \in (0,1)$ we model the shape of the quantiles as a function of $d$ using a quantile regression with a truncated Fourier series approach
\begin{align*}
			d \mapsto \tilde{G}_{d,h}^{\tau}=\beta_{0,h}+\sum_{i=1}^p\beta_{i,h}\cos\left(\frac{2\pi d i}{365.25}\right)+\sum_{j=1}^q\beta_{p+j,h}\sin\left(\frac{2\pi d j}{365.25}\right)+\beta_{p+q+1,h}g^{toa}_{d.h}.
		\end{align*}
We use truncated Fourier series with the TOA as an extra exogenous input factor, as it captures the seasonalities of the upper bound quite good. Truncated Fourier series are used frequently in the literature to model the seasonality of meteorological variables like e.g. in \cite{stoll2010}. We use information criteria like the \textit{Bayesian information criterion} (BIC) to choose
 the right number of sine and cosine terms.  The BIC suggests $p=q=2$. With this choice both yearly and half yearly trigonometric effects are taken into account. The parameter estimation for the quantile regression can be formulated as a linear programming problem and solved with any LP solver. For this purpose and for all other model building results in this paper as well as graphics we use the software package MATLAB, provided by MathWorks.

With a look at the so-called mean residual life plot we choose the quantile $\tau=0.75$ as a reasonable threshold for POT. This choice ensures on the one hand that the Pickands-Balkema-de Haan theorem holds, which states that the distribution of the exceedances can be approximated well by a generalized Pareto distribution (GPD). On the other hand the choice ensures that there are enough exceedances for parameter estimation of the GPD. We compute the threshold excess
		\begin{align*}
			u_{d,h}^{0.75}=\tilde{g}_{d,h}^+-\tilde{G}_{d,h}^{0.75}
		\end{align*}	
for the time varying threshold $\tilde{G}_{d,h}^{0.75}$ and the exceedances $\{\tilde{g}_{d,h}^+:\tilde{g}_{d,h}^+ >\tilde{G}_{d,h}^{0.75}\}$. 
It is not surprising that we still observe some seasonality in these absolute exceedances, as they are on average higher around noon compared to the morning or late evening. For the natural alternative of relative
exceedances, however, we also observe a similar behavior in the other direction. They have on average relatively high values in the morning and in the evening due to the low irradiation then. 
Therefore we work with the absolute exceedances and deseasonalize them to get

		\begin{align*}
			\tilde{u}_{d,h}^{0.75}=\frac{u_{d,h}^{0.75}}{E_h},\quad E_h>0,
		\end{align*}	
where $E_h$ is the average exceedance over the threshold in hour $h$ estimated using a linear regression with truncated Fourier series:
\begin{align*}
\label{Fourier}
	E_h=\beta_0+\sum_{i=1}^p\beta_{i}\cos\left(\frac{2\pi h i}{24}\right)+\sum_{j=1}^q\beta_{p+j}\sin\left(\frac{2\pi h j}{24}\right).
\end{align*}
Here again, the BIC suggests $p=q=2$. The adjusted threshold excesses $\tilde{u}_{d,h}^{0.75}$ can now be considered as a stationary time series. So we are able to fit a generalized Pareto distribution to the deseasonalized threshold excesses. The parameter estimation is done with maximum-likelihood estimation. As the estimated shape parameter $\xi$ is negative the GPD has an upper endpoint $r_u$. We compute the right endpoint $r_u=-\frac{\sigma}{\xi}$ with the scale parameter $\sigma$ and the shape parameter $\xi$ of the GPD.  We get as a final estimator of the upper bound for every day $d$ and hour $h$ the value
		\begin{align*}
			g_{d,h}^+=\tilde{G}_{d,h}^{0.75} + E_h r_u.
		\end{align*}
Next consider the lower bound of GHI. The steps of the estimation procedure for the lower bound are quite similar. In the first step historical minimum irradiation values 
\begin{align*}
	\tilde{g}_{d,h}^-=\min_{i=1,...,n}g_{d,h}^{(i)},\quad d=1,\dotsc,365, \ h=0,\dotsc,23,
\end{align*}
are computed for the years in the learn period. However, instead of using these values for a quantile regression with a low quantile we consider the relative deviations from the estimated upper bound caused by clouds and transform it with a strictly monotone link function $\ell:(0,1)\rightarrow \mathbb{R}$ to get
\begin{align*}
	c_{d,h}=\ell\left( \frac{g_{d,h}^+-\tilde{g}_{d,h}^-}{g_{d,h}^+}\right),
\end{align*}
where we use the logit-link $\ell(x)=\log(x/(1-x))$.
The deviations caused by clouds $c_{d,h}$ are similar to the solar lost component mentioned in \cite{safi2002}.  
With a link function we prevent that $g_{d,h}^-$ takes values lower than the natural bound zero. An overview of further possible transformation functions is given in \cite{atkinson1985}.

As before a quantile regression with $\tau=0.75$ is performed with the same choice of hourly $p$ and $q$ as in the case of the upper bound to get $\tilde{C}_{d,h}^{0.75}$. In contrast to the upper bound the threshold excesses here show no seasonalities in the course of the day and are directly fitted with a generalized Pareto distribution. With the right endpoint $r_\ell$ we obtain an estimation of the lower bounds depending on the estimated upper bound for every day $d$ and hour $h$ as
\begin{align*}
	g_{d,h}^-=g_{d,h}^+\cdot(1-\ell^{-1}(\tilde{C}_{d,h}^{0.75}+r_\ell)).
\end{align*}
For the estimated bounds we get
\begin{align*}
	0<g_{d,h}^-<G_{d,h}<g_{d,h}^+<g^{toa}_{d,h}
\end{align*}
for all $d$ and $h$ with $\mathbb{E}(G_{d,h})>0$. The only exception is in the case of sunrise and sunset hours. In this case the lower bound is not estimated with the algorithm above but set to a value of zero to simplify the estimation process. Plots of the upper and lower hourly bounds together with historical observations for each first day of a month are shown in figure 3.1. Seasonalities both on daily and hourly levels are clearly visible. The upper bound is much closer to the observations than the TOA (compare also figure 2.1). 
\begin{figure}[h]
	\begin{center}
		\includegraphics[width=\textwidth]{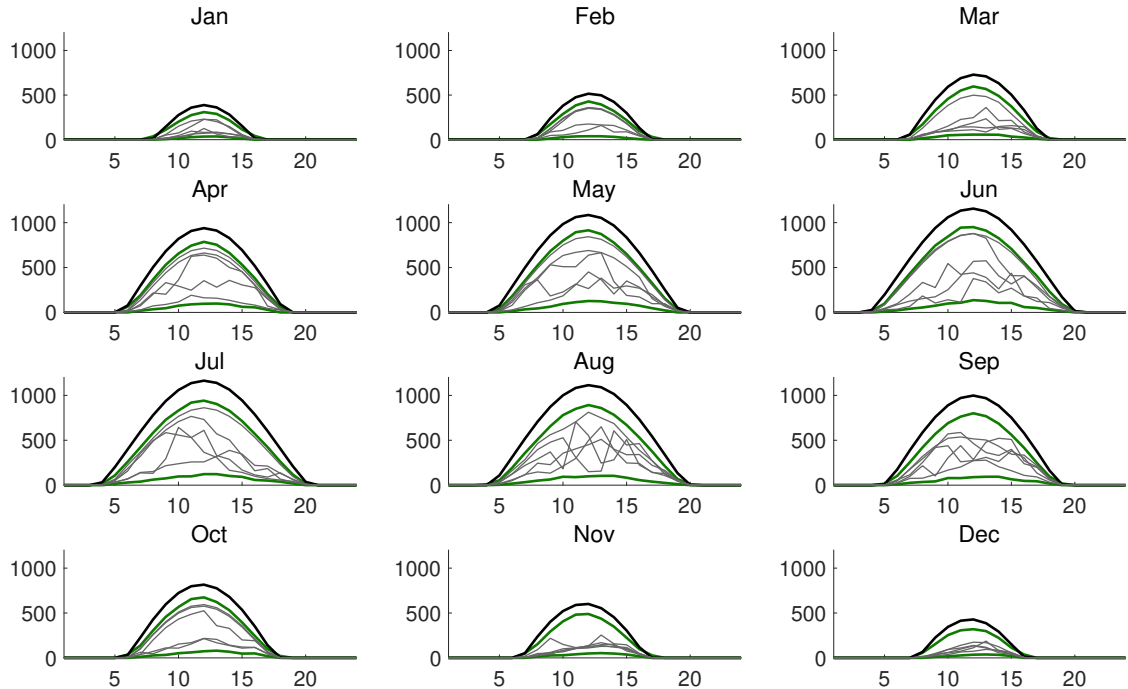}
		\caption[Estimated lower and upper bounds.]{Estimated lower and upper bounds for the first day of every month with observations from 5 years. Hourly values for lower and upper bound as well as hourly GHI in $\mathrm{Wh/m^2}$.}
		\label{boundsYear2}
	\end{center}
\end{figure}

\subsection{Hourly marginal distributions}
After deriving estimators $g_{d,h}^-$ and $g_{d,h}^+$ for the bounds of the GHI $G_{d,h}$ we will now fit a marginal distributions to the hourly sun intensities with a suitable time dependent distribution. The hourly intensities of the sun 
$$
M_{d,h}=(G_{d,h}-g_{d,h}^-)/(g_{d,h}^+-g_{d,h}^-)
$$ 
now assume values in the interval (0,1). A distribution that is very flexible for modeling continuous data on the interval (0,1) is the beta distribution. By variation of its two parameters its density can take a variety of different shapes. \cite{engeland2017} describe that the hourly clearness index has an unimodal asymmetric behavior. If one regards the clearness index on a shorter period (e.g minutes) one observes a bimodal distribution caused by the cloud effects. The beta distribution is capable to describe both cases of unimodal as well as bimodal continuous distributions on $(0,1)$ depending on its parameters. Already \cite{graham1990} used the beta distribution to model the distribution of the clearness index. As the sun intensity is comparable to the clearness index we therefore consider the beta distribution as a good choice for the hourly marginal distributions of the sun intensity, too. We will include seasonality of its parameters using a beta regression. \cite{ferrari2004} present the concept of a beta regression and introduce an alternative parametrization of the beta distribution that we use here, too. With a mean parameter $\mu$ and a precision parameter $\phi$ the density of the beta distribution $\mathrm{Beta}(\mu,\phi)$ is of the form:
\begin{align*}
	f(x)=\frac{\Gamma(\phi)}{\Gamma(\mu\phi)\Gamma((1-\mu)\phi)}x^{\mu\phi-1}(1-x)^{(1-\mu)\phi-1},\quad 0 < x < 1,
\end{align*} 
where $\Gamma(\cdot)$ is the gamma function. A random variable $X$ with this distribution has mean $E(X) = \mu$ and variance $V(X) = \mu(1-\mu)/(1+\phi)$. Therefore $\phi$ can be considered as a
precision parameter as the variance decreases with increasing $\phi$. 

We now assume that the hourly intensities $M_{d,h}$ are distributed as 
\begin{align*}
	M_{d,h}\sim \mathrm{Beta}(\mu_{d,h},\phi_{d,h})
\end{align*}
with time dependent mean and precision parameters. As shown in \cite{ferrari2004} we express the mean $\mu_{d,h}$ and the dispersion $\phi_{d,h}$ through the functional relationships. The parameters are transformed with appropriate link functions $\ell_1$ and $\ell_2$, respectively. We denote by $\Lambda_{d,h}=\mathbb{E}(G_{d,h})$ the expected irradiation at day $d$ and hour $h$ that again is estimated with a linear regression based on truncated Fourier series as shown in section 2.
The mean  parameters are then expressed in the form
\begin{align*}
	\ell_1(\mu_{d,h})=\zeta_{h,1}+\zeta_{h,2} \Lambda_{d,h}
\end{align*} 
with the logit link function $\ell_1$ and the unknown parameters $\zeta_{h,1}$ and $\zeta_{h,2}$. Similarly in the case of the precision parameter:
\begin{align*}
	\ell_2(\phi_{d,h})=\vartheta_{h,1} + \vartheta_{h,2} \Lambda_{d,h},
\end{align*}
where the logarithm is used as the link function $\ell_2$ and $\vartheta_{h,1}$ and $\vartheta_{h,2}$ are the unknown parameters. 

As an alternative to the beta regression one could also perform a quantile regression for every hour with a sufficient number of quantiles as a nonparametric alternative to obtain empirical distribution functions. \cite{golastaneh2017} use a quantile regression with weather variables as exogenous input to estimate empirical distributions of hourly PV yields. In this paper we prefer the parametric approach using a beta distribution as a marginal distribution as it is better suited for a combination with a copula approach to describe the dependence structure as suggested in the next subsection.

\subsection{Copula-based time series model}
After the treatment of hourly marginals we now have a look on the dependence structure of the time series $(M_{d,h})$. In time series analysis one very often uses Gaussian processes like ARMA processes
after transforming the data such that one can assume the marginal distributions to be Gaussian, see \cite{box2008}. However, this approach has serious drawbacks in this case, as Gaussian processes always have the property of tail independence as we will describe below. Therefore we will take hourly and daily dependencies into account with a copula-based univariate time series model as described in \cite{patton2012}. Let us first consider the model for bivariate dependencies. The marginal distributions for hourly irradiation have already been determined in the last section. Therefore it is a natural approach to look at the copulas that couple the marginal distributions to get multivariate distributions, see \cite{nelsen2006}, \cite{joe1997} or \cite{joe2014} for detailed treatments of the theory of copulas. 

We will repeat here the most important definitions and properties of bivariate copulas. For two random variables $X$ and $Y$ with marginal distributions $F(x) = \mathbb{P}(X \le x)$ and $G(y) = \mathbb{P}(Y \le y)$, $x,y \in \mathbb{R}$, the joint distribution function $H$ can be written in the form
$$
H(x,y) = \mathbb{P}(X\le x,Y\le y) = C(F(x),G(y)), \quad x,y \in \mathbb{R},
$$
where the function $C:[0,1]^2 \to [0,1]$ is called the \textit{copula}. In the case of continuous random variables we can use the probability integral transform (PIT) to transform $X$ and $Y$ to standard uniform random variables $U$ and $V$ via $U = F(X)$ and $V = G(Y)$. The copula is then just the joint distribution function of $(U,V)$, i.e.
$$
H(u,v) = \mathbb{P}(U \le u,V \le v), \quad 0 < u,v < 1.
$$
An important descriptive measure of the dependence is given by the so-called \textit{quantile dependence function} $\lambda^q, \ 0 < q < 1$, defined as follows (see Patton (2013)):
\begin{align*}
	\lambda^q&=\begin{cases}
		\mathbb{P}(U\leq q|V\leq q) = \frac{C(q,q)}{q},\quad 0<q\leq 0.5,\\[3mm]
		\mathbb{P}(U> q|V> q) = \frac{1-2q+C(q,q)}{1-q},\quad 0.5<q<1.
	\end{cases} 
\end{align*}	
Given empirical observations $(u_1,v_1), \dotsc, (u_N,v_N))$ that are often given in terms of relative ranks, one gets the empirical version
\begin{align*}	
	\tilde{\lambda}^q&=\begin{cases}
		\frac{1}{Nq}\sum_{i=1}^N \mathds{1}_{\{u_{i}\leq q, v_{i} \leq q\}},\quad &0<q\leq 0.5\\
		\frac{1}{N(1-q)}\sum_{i=1}^N \mathds{1}_{\{u_{i}>q,v_{i}>q\}},\quad &0.5<q<1.
	\end{cases}
\end{align*}
Taking the limits $q \to 0$ and $q \to 1$ we get the \textit{upper tail dependence coefficient} $\lambda_U$ and the \textit{lower tail dependence coefficient} $\lambda_L$  as 
\begin{align*}
	\lambda_U:&=\lim_{q\rightarrow 1}\mathbb{P}(U>q|V>q) = \lim_{q\rightarrow 1} \frac{1-2q+C(q,q)}{1-q}
\end{align*}
and
\begin{align*}
	\lambda_L:&=\lim_{q\rightarrow 0}\mathbb{P}(U<q|V<q) = \lim_{q\rightarrow 0} \frac{C(q,q)}{q}.
\end{align*}

Empirical tail dependence coefficients are presented in table 3.1 for hours between 9 am and 3 pm. They are estimated with the methods described in Patton (2013), section 3.4.1. We clearly see that the tail dependence coefficients are significantly positive and tail dependence between consecutive hours is the highest around noon and decreases in the directions to the morning and evening. It is not surprising that with increasing lags between the hours tail dependence decreases. We also see that the upper tail dependence is stronger than the lower tail dependence. These properties are quite intuitive. The higher upper tail dependence around noon can be justified with the higher share of direct irradiation that is dependent on the cloud conditions.

\begin{table}[h]
\centering
	\captionabove[Empirical tail dependence]{Overview of empirical tail dependence for pairs of hours between 9am and 3pm. Empirical upper tail dependence (right upper triangle) and empirical lower tail dependence (left lower triangle).}
	\label{t:tailDependence}
	\begin{tabular}{|	l	||	lllllll	|}
	\hline
		&\multicolumn{1}{c}{9}&\multicolumn{1}{c}{10}&\multicolumn{1}{c}{11}&\multicolumn{1}{c}{12}&\multicolumn{1}{c}{13}&\multicolumn{1}{c}{14}&\multicolumn{1}{c|}{15}\\
	\hline
	\hline
	 9&\multicolumn{1}{c}{-}&        0.674&        0.559&        0.507&        0.538&        0.463&        0.464\\
        10&0.646&            \multicolumn{1}{c}{-}&        0.782&        0.694&        0.654&        0.535 &       0.414\\
        11&0.522&        0.676  &          \multicolumn{1}{c}{-}&        0.827&        0.716&        0.594&          0.4\\
         12&0.33&        0.459  &      0.599  &          \multicolumn{1}{c}{-}&        0.808&         0.64&        0.434\\
        13&0.267&        0.395 &       0.491&        0.681&            \multicolumn{1}{c}{-} &       0.732&        0.511\\
        14&0.264&        0.365 &       0.451&        0.535 &       0.641&           \multicolumn{1}{c}{-} &        0.67\\
        15&0.223&        0.303&         0.38 &       0.427&          0.5 &       0.607&            \multicolumn{1}{c|}{-}\\
	\hline	
	\end{tabular}
\end{table}

A classical time series approach based on Gaussian processes would imply that all bivariate distributions have a Gaussian copula with the property $\lambda_L = \lambda_U = 0$. This is clearly not the case here. Therefore we look for other time series models based on the copula approach. As the dependence decreases with lag, a first order Markov process approach based on copulas as described in Joe (1997), section 8, is a natural candidate to describe the dependence structure within a day.

Here the bivariate vector of consecutive hourly intensities is assumed to be distributed like a time dependent distribution function $H_{d,h}$. We make the simplifying assumption
that the parameters of this copula only depend on the hour $h$ but not the day $d$ as we could not find any structural relation between the day $d$ and the structure of this bivariate copula. 
According to Sklar's theorem we can thus decompose the bivariate distribution into a copula and its univariate marginals.
\begin{align*}
	\mathbb{P}(M_{d,h} \le x,M_{d,h+1} \le y)&=C_{h}(F_{d,h}(x),F_{d,h+1}(y)).
\end{align*}
The hourly marginal distributions $F_{d,h}$ are assumed to follow time dependent beta distributions as discussed in the previous subsection. 
With the estimated parameters of the time dependent beta distributions we use the probability integral transform (PIT) to get
\begin{align*}
	U_{d,h}=F_{d,h}(M_{d,h}).
\end{align*}
The PIT does not lead to perfect uniform distributions for $U_{d,h}$ as we use an estimated beta distribution for the transformation. Therefore we consider the ranks of $U_{d,h}$ to estimate the copula. The inference based on ranks is considered in \cite{oakes1982}. The type of inference that we use here is also related to the method of \textit{inference functions for margins} as described in Joe (1997).

\begin{figure}
	\begin{center}
		\includegraphics[width=\textwidth]{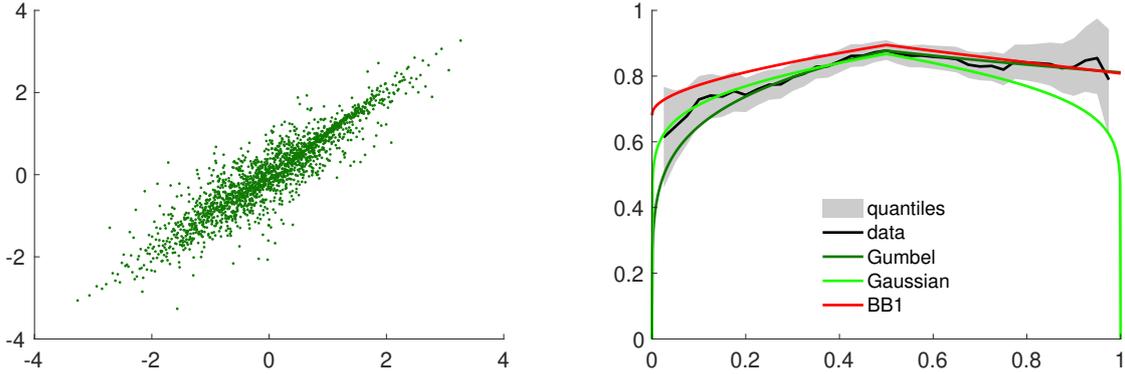}
	\caption[Quantile dependence plots.]{Scatterplot and quantile dependence plot for the empirical copula around noon. The empirical upper and lower tail dependence is represented by the red stars and the green star represents the upper tail dependence of the Gumbel copula.}
		\label{scatterQuantile}
	\end{center}
\end{figure}
As proposed in \cite{patton2013} we use so-called quantile dependence plots to compare the quantile dependence of different copulas to real data. In our case, there is a strong upper tail dependence that means that hours with a high intensity of the sun are often followed by another hour with a high intensity, too. To provide an example we show a scatter plot of the 12am and 1pm hours and the corresponding quantile dependence in figure \ref{scatterQuantile}. For the scatter plot we transformed the marginal distributions to standard normal distributions as this gives a more informative picture. This also shows that a Gaussian copula is probably not the most appropriate model as that would imply an elliptical shape of this scatter plot. In the right plot the quantile dependence plot for the observed data and for some copulas is presented for $q\in[0.025,0.975]$. The grey shaded area is a 90\% confidence band. Plots for other hours look similar with a decreasing upper tail dependence going to the morning or evening hours.  

In this paper we consider the following examples of bivariate copulas. 

\begin{itemize}
\item As a first example we consider the Gaussian copula that is obtained as the copula of a bivariate normal distribution with possible correlation.
The Gaussian copula is used as a kind of benchmark as it is often used in time series modelling.
It has the drawback that its tail dependence coefficients always fulfill $\lambda_L = \lambda_U = 0$, even for very high correlation coefficients, see \cite{nelsen2006}. 

\item As an example with the 
characteristic of upper tail dependence $\lambda_U > 0$ we choose the so-called Gumbel copula, which is an example from the family of Archimedean copulas having a functional form 
$C(u,v)=\psi (\psi^{-1}(u)+\psi^{-1}(v))$ with a so-called generator $\psi:[0,\infty)\rightarrow [0,1]$. 
The Gumbel copula has the generator $\psi (t) = \exp(-t^{1/\theta}), \ t \ge 0,$ with parameter $\theta>1$ and tail dependence coefficients $\lambda_U=2-2^{1/\theta}$ and $\lambda_L = 0$.  

\item Besides the Gumbel copula we also consider the BB1 copula (\cite{joe2014}) as a standard example of a copula with arbitrary upper and lower tail dependence. It is the Archimedean copula with generator is given as $\psi(t)=(1+t^{1/\delta})^{-1/\theta}, \ t\geq 0,$ with two parameters $\theta >0, \delta >1$. There is a direct link between its parameters and the tail dependence coefficients: $\lambda_L=2^{-1/(\delta\theta)}$ and $\lambda_U=2-2^{1/\delta}$.
\end{itemize}

\begin{table}[h]
\centering
	\captionabove[Tail dependence]{Overview of empirical lower and upper tail dependences of the other locations for the 12pm/1pm dependence.}
	\label{t:tailDependence_all}
	\begin{tabular}{|	l	||	llll	|}
	\hline
		&Athens&Nairobi&Reykjavik&Siegen\\	
	\hline
	\hline
		$\lambda_L$&0.597&0.455&0.556&0.681\\
		$\lambda_U$&0.835&0.665&0.688&0.808\\
	\hline
\end{tabular}
\end{table}

\begin{figure}
	\begin{center}
		\includegraphics[width=\textwidth]{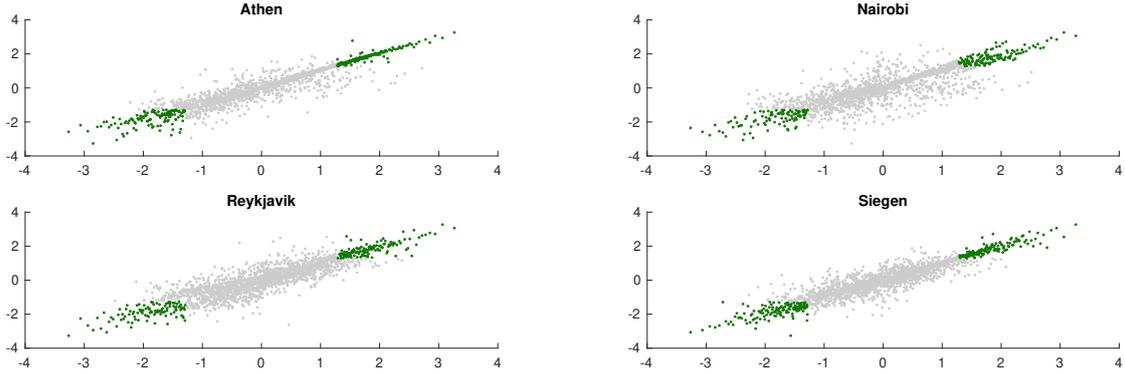}
	\caption{Scatterplot of all locations.}
		\label{scatter_all}
	\end{center}
\end{figure}

Before we move on we want to take a look at the dependence structure, especially the lower and upper tail dependence, of the three other locations mentioned in section 2. Table \ref{t:tailDependence_all} shows the lower and upper tail dependence of Athens, Nairobi and Reykjavik and figure \ref{scatter_all} provides examples of scatterplots of the 12am and 1pm hour for all considered locations. We see that the tail dependence differs from location to location with the highest upper tail dependence in Athens and the lowest upper tail dependence in Nairobi. Depending on the location under consideration the correct recognition of the existence of tail dependence becomes more or less important.

We have to describe now how to get the general dependence of the whole time series. A very popular technique to derive multivariate distributions from bivariate ones is given by the method of vine copulas 
introduced by \cite{bedford2002}, see also \cite{czado2010} and \cite{niko2012}. They are based on the idea of glueing together bivariate copulas to so-called Markov trees and then adding a hierarchy of additional layers of conditional dependence. 
Here we only consider the simple case of only one layer of dependence described by a Markov tree as in section 2 of \cite{bedford2002}. 

As a simple first model we consider the case that irradiation on different days is stochastically independent and the dependence within a day is described by a classical Markov chain, which is determined by a sequence of given 
bivariate copulas $C_h$. 
As the dependence varies in the course of the day it is clear that one copula with a fixed parameter for all hours of the day is not suitable. Dependence around noon would be underestimated while the dependence in the morning would be overestimated. Therefore we estimate copulas with different parameters depending on the hour $h$ of the day. For this we use maximum pseudolikelihood estimation in the case of Gaussian and Gumbel copula. For a detailed description of the method of maximum pseudo likelihood estimation we refer to \cite{genest1995}. We use a transformation with the estimated beta distribution and maximize a rank-based log likelihood. For the parameters of the BB1 copula we use the empirical tail dependence coefficients obtained by tail copulas and use the link between the parameters and the tail dependence. Yearly seasonal effects in the parameters are not observed. One advantage of the copula-based univariate time series model over a classical time series model like an ARMA($p,q$)-process lies in the characteristic that there are clusterings of exceedances of high thresholds. This is taken into account through the positive upper tail dependence coefficient of the Gumbel and BB1 copula.

We also consider a second a bit more sophisticated approach where we take both daily and hourly dependencies into account. For this purpose we add another bivariate copula $C^*$
that describes the dependence between day $d$ and the consecutive day $d+1$. If we would use a classical Markov chain approach, this copula $C^*$ would describe the dependence
between the last hour $h_2$ of the evening of day $d$ and the first hour $h_1$ in the morning of day $d+1$. However, as irradiation is very low in the morning and evening, this approach does not work well.
Correlations and tail dependence between the same hours of consecutive days are the highest for the noon hour, which is the hour with the highest expected hourly irradiation. We observe a clearly asymmetric tail dependence with a upper tail dependence, that is a bit lower ($\tilde{\lambda}_U^{12}=0.4$) compared to the intraday copulas. There is no lower tail dependence visible in the empirical quantile dependence function, see figure 3.3.

\begin{figure}[h]
	\begin{center}
		\includegraphics[width=\textwidth]{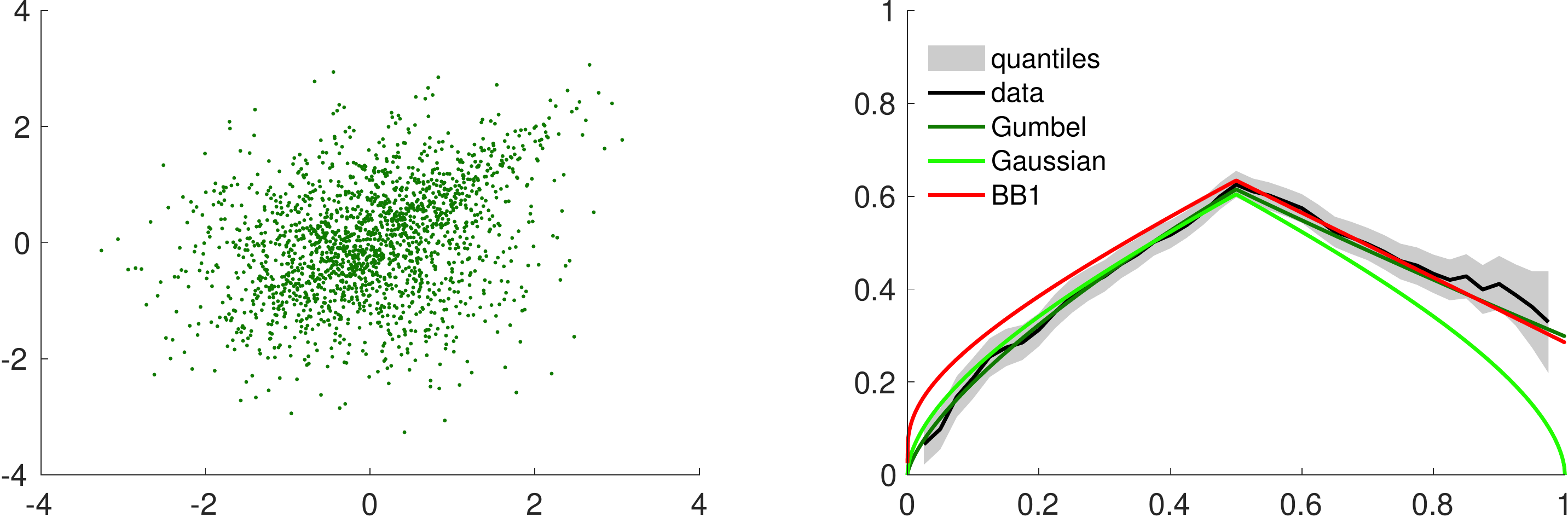}
	\end{center}
	\caption[Copula for joint model.]{Scatterplot and quantile dependence plot for the empirical noon copula $C^*$.}
	\label{copulaIrradiationPV1}
\end{figure}
\begin{figure}[h]
	\begin{center}
		\includegraphics[width=\textwidth]{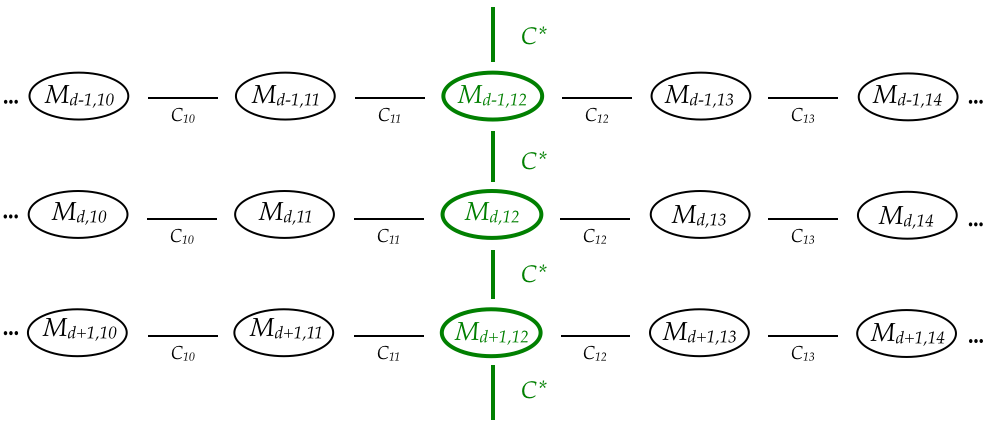}
	\end{center}
	\caption[Visualization of the second copula approach]{Tree structure of the second copula approach.}
	\label{copula12}
\end{figure}

We now consider a Markov sequence of bivariate copulas $C^*$ to include the daily dependence structure in the noon hour. The joint distribution function of two consecutive noon hours is then given as
\begin{align*}
	\mathbb{P}(M_{d,12} \le x,M_{d+1,12} \le y)&=C^*(F_{d,12}(x),F_{d+1,12}(y)).
\end{align*}
Thus we can define a Markov chain structure for the irradiation at the noon hour. The intraday dependence is included with the same approach as before. Taking into account
both kinds of dependence we get a Markov tree as described in figure \ref{copula12}. 

Like in the case of hourly time dependencies we use Gumbel, Gaussian and BB1 copulas for the daily dependencies described by $C^*$.

We can now generate sequences of hourly irradiations for one year with the algorithm described below. If the copula $C_h$ is the joint distribution function of a random vector $(U,V)$, we can derive the conditional
distribution function 
\begin{align*}
	c_{h|u}(v)= \mathbb{P}(V\le v|U=u) = \frac{\partial C_h(u,v)}{\partial u}. 
\end{align*}
We denote by $c_{h|u}^{-1}$ the inverse of $c_{h|u}(v)=\partial C_h(u,v)/\partial u$ and $c_{u}^{-1}$ shall denote the inverse of $c^*_{u}(v)=\partial C^*(u,v)/\partial u$ (for a simplification of the notation we omit here the * in the notation for the inverse).
\begin{enumerate}
	\item Generate independent uniform (0,1) variates $v_1,v_2,\dotsc,v_{365}$.
	\item Set $u_1^* = v_1$ and  $u_d^*=c_{u_{d-1}}^{-1}(v_d)$ for $d=2,3,\dotsc,365$.
	\item For each fixed day $d=1,2,3,\dotsc,365$ do the following:
	\begin{enumerate}
	\item Determine the first and the last hour $h_1$ respectively $h_2$ with positive expected irradiation.
	\item Generate independent uniform (0,1) variates $v_{h_1},v_{h_1+1}..,v_{11},v_{13},\dotsc,v_{h_2}$.
	\item Set $u_{12}=u_d^*$.
	\item For $j=11,10,\dotsc,h_1$ set $u_j=c_{h|u_{j+1}}^{-1}(v_j)$.
	\item For $j=13,14,\dotsc,h_2$ set $u_j=c_{h|u_{j-1}}^{-1}(v_j)$.
	\item For all $h=h_1,\dotsc,h_2$ we obtain simulations of global horizontal irradiation:
	\begin{align*}
		G_{d,h}=F_{d,h}^{-1}(u_h)\cdot (g_{d,h}^+-g_{d,h}^-)+g_{d,h}^-.
	\end{align*}
		\end{enumerate}
\end{enumerate}
Using this algorithm we get daily vectors of hourly irradiation $\mathbf{G}_d$ for a whole year.

\subsection{Evaluation methods}
In this article we consider probabilistic forecasts of hourly GHI. We want to compare our models to other benchmark models. Besides the evaluation of the hourly marginals the evaluation of the hourly dependence structure is of great importance. For the comparison of probabilistic forecasts and realizations we use so-called scoring rules. A scoring rule is a function that assigns a numerical score to a predictive distribution $F$ and an observation $x$. It can be applied both to univariate and multivariate distributions. We refer to \cite{gneiting2012} and \cite{gneiting2014} for further information on the characteristics of scoring rules and probabilistic forecasts in general.

We use various scoring rules that emphasize different characteristics of the models. To evaluate marginal distributions the continuous ranked probability score (CRPS) is a frequently used scoring rule. The CRPS with the quantile decomposition (\cite{laio2007}) is given by
\begin{align*}
	\mathrm{CRPS}(F,x)&=\int_0^1 2(\mathds{1}_{\{x<F^{-1}(\tau)\}}-\tau)(F^{-1}(\tau)-x)d\tau.
\end{align*}
In most studies that compare different models for probabilistic forecasting, this scoring rule or variants of it that only consider a finite number of quantiles are used, see e.g. \cite{hong2019}. 
As the CRPS considers only marginal distributions it is not able to detect misspecifications in the multivariate dependence structure in its original formulation. One can use functionals that map a multivariate term to a univariate one to evaluate the dependence structure in a roundabout way. We consider $X=\kappa (\mathbf{G}_d)\sim F$ with a functional $\kappa$. One functional that we consider in this paper is similar to an event that is defined in \cite{pinson2012} for the evaluation of wind forecasts with event based scores. They consider the event, that the wind power is higher than a threshold for a period of six consecutive hours. This event is also useful in the case of irradiations or PV energy. We extend the idea and use the sum of the exceedances above a threshold that is 80\% of the estimated upper bound of the hours between 10am and 3pm conditional on the event that the irradiation is higher for the whole period. We choose this time period as the irradiation is the highest around noon. Thus, we obtain the functional
\begin{align*}
	X_1 = \kappa_1(\mathbf{G}_d)=\left(\sum_{h=10}^{15}G_{d,h}\right)\cdot\left(\prod_{h=10}^{15}\mathds{1}_{\{G_{d,h}>0.8g_{d,h}^+\}}\right).
\end{align*}
This functional is important in the optimal sizing of a solar accumulator where one is interested in the sum of PV yields above a threshold for consecutive hours. Here, the functional is useful in detecting misspecifications in the upper tail dependence. When we compute the CRPS of the distribution of this quantity $X_1 = \kappa_1(\mathbf{G}_d)$ then we will call this CPRS-U. As another example of such a functional we will compute the weekly sum
$$
X_2 = \kappa_2(\mathbf{G}_d,\ldots,\mathbf{G}_{d+6}) = \sum_{j=d}^{d+6} \sum_{h=1}^{24} G_{j,h}.
$$
The CRPS of this quantity will be denoted as CRPS-W.
\\
The energy score (see \cite{gneiting2012}) as the multivariate generalization of the CRPS is a scoring rule that is frequently used to evaluate multivariate models. It can be written as
\begin{align*}
	\mathrm{ES}(F,\mathbf{x})&=\mathds{E}_F||\mathbf{X}-\mathbf{x}||-\frac{1}{2}\mathds{E}_F||\mathbf{X}-\mathbf{X}'||
\end{align*}
with independent copies $\mathbf{X}$ and $\mathbf{X}'$ of a random vector with distribution $F$.
A further scoring rule that we consider is the variogram-based score (\cite{scheuerer2015}) as a third scoring rule for the hourly dependencies. It is based on pairwise differences and proposed to evaluate multivariate forecasts. For a given $n$-variate observation vector $\mathbf{x}=(x_1,\dotsc,x_n)^T$ and a multivariate forecast distribution $F$ the variogram score of order one is defined as
\begin{align*}
	\mathrm{VS}(F,\mathbf{x})=\sum_{i,j=1}^n \omega_{ij}\left( |x_i-x_j| - \mathds{E}_{F}|X_i-X_j| \right)^2.
\end{align*}
Here, $X_i$ and $X_j$ are components of a random vector $\mathbf{X}=(X_1,\dotsc,X_n)^T$ with distribution function $F$. The nonnegative weights $w_{ij}$ are used to emphasize or downweight certain pairs of squared variogram differences. We choose $w_{ij}=c_{ij}$, where $c_{ij}$ is the Pearson correlation coefficient for hours $i$ and $j$, so that the pairs of squared variogram differences of consecutive hours are more emphasized. An approximation of the forecast variogram $\mathds{E}_F|X_i-X_j|$ is given by
\begin{align*}
	\mathds{E}_{F}|X_i-X_j| \approx \frac{1}{m}\sum_{k=1}^m |x_i^{(k)}-x_j^{(k)}|\quad , i,j=1,...n
\end{align*}
with the forecast distribution expressed in the form of an ensemble $\mathbf{x}^{(1)},\dotsc,\mathbf{x}^{(m)}$. As the irradiation is highest around noon it is important to have a reliable model for the dependence structure around noon. Therefore we consider the variogram score for the six hours around noon ($h=10,\dotsc,15$). This matches with the choice of hours for the functional $\kappa_1$.

A drawback of the variogram score is that a bias that is the same in all components cannot be detected. Therefore the variogram score should always be used together with scoring rules like the CRPS that evaluates the marginal distributions. With these procedures the forecasting ability of probabilistic models can be evaluated with regard to their dependence structure as well as marginal distributions.

One important question in the evaluation of different models is whether the differences between the scores of the models are significant. For this purpose we use the Diebold-Mariano test (see \cite{diebold1995}). The hypothesis of equal forecast accuracy of two models is tested against its alternative for arbitrary loss functions, that are in our case the introduces scoring rules.

\section{Empirical results}
\begin{figure}[h]
	\begin{center}
		\includegraphics[width=\textwidth]{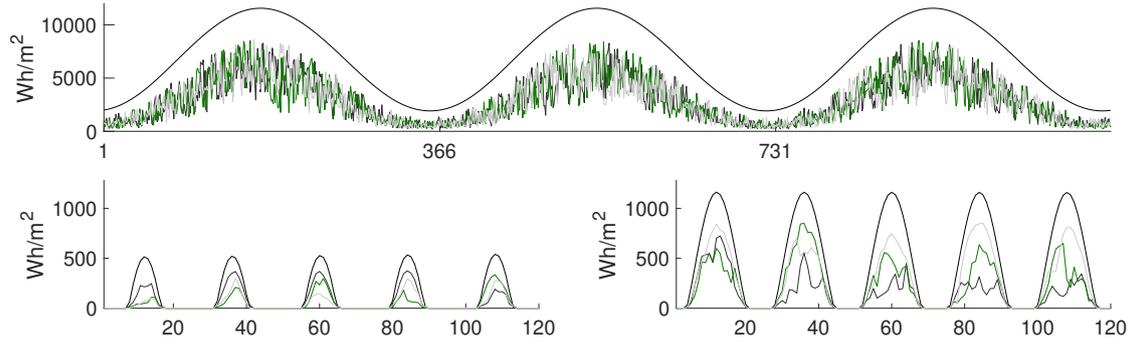}
	\end{center}
	\caption{Simulation paths for three years of daily GHI (above) and five days of hourly GHI in winter (bottom left) respectively summer (bottom right) with TOA.}
	\label{simulationsC}
\end{figure}

In this section we present empirical results for our models. First, we evaluate the copula-based time series models with the scoring rules that were introduced in the last section. We also compare the estimated bounds with alternative bounds that are frequently used or naturally given by the TOA with weighted scoring rules.\\
Daily and hourly simulations of the copula-based time series model with the second approach and a Gumbel copula are illustrated in figure \ref{simulationsC}. Compared to the historical observations of daily and hourly GHI in figure \ref{irradiation} we see that our model fulfills the stylized facts mentioned in section 2 quite well. By visual inspection we see that the seasonalities as well as daily and hourly dependencies are well mapped by the model.
\subsection{Evaluation results - univariate copula-based models}
We use a Monte-Carlo approach and generate 10000 sequences of hourly GHI of each one year with the proposed copula-based time series model. For every model we obtain empirical cumulative distribution functions of 10000 values for every hour. To obtain the CRPS we proceed like it is suggested in the R package \textit{scoringRules} described in \cite{jordan2018}. We compute the empirical quantiles with $\tau\in\{0.001,\dotsc,0.999\}$ for every hour and calculate the empirical CRPS with the quantile decomposition. To obtain one single score for every model we take the average over all hourly scores. Similarly, the CRPS with the introduced functional $\kappa$ is averaged over all times. Furthermore, we also calculate the CRPS of the cumulated weekly irradiations to take the daily correlations into account. The scores based on the variogram score are also averaged to get one single score.

The absolute value of a score of one single model does not have a meaning in itself. Therefore they have to be compared to the scores of easy benchmark models. In this article we use two benchmark models. The first one is a classical historical simulation (HS). We randomly draw one of the 7 values from the same day and hour from one of the 7 years in the learn data set described below. The second model uses a deterministic allocation (DA) of the daily irradiation to the hours of the day that is similar to the approach used by \cite{wagner2014}. In his article he uses a daily pattern transformation that can be for example step functions with one step for each hour or appropriately scaled Gaussian functions. The shape of the intraday pattern is similar to the estimated upper bound. 

Our stochastic models based on the different approaches with the copula-based time series models are referred to as C1 (different days are assumed to be stochastically independent) and C2 (daily dependencies given by the Markov tree structure). We use cumulative daily irradiations that we obtain from the C2-Gumbel model in the second benchmark model (DA). The resulting scores for the two benchmarks and the stochastic models are shown in table 4.3. There the used copula is written after the abbreviation for the variant. We compute the CRPS for every hour of the year with positive irradiation and take the average score (CRPS-H). We proceed analogously with the weekly CRPS (CRPS-W) and the energy score (ES). For the variogram score (VS) we use the weights given by the corresponding correlations. The last score that we consider is the CRPS derived with the functional $\kappa_1$ with a special emphasize of consecutive values above a threshold, that can detect misspecifications in the upper tail dependence (CRPS-U). For ES and VS we use the 7-dimensional data of the main hours 10 to 16 that are also used as a basis of CRPS-U. We normalize the scores such that the simplest benchmark model HS always assumes the value 1.

The historical database is divided into a learn period of 7 years ranging from 2005 to 2011 for calibrating our model and a test period ranging from 2012 to 2018 for the out of sample evaluation. We consider it as important to cover a few years in the learn as well as in the test period as there are always differences in the characteristics of various years like especially sunny years. For example the empirical upper tail dependence between the 12am and 1pm hour for five different years ranges between 0.6 and 0.85. With a reasonable diversity it can be prevented that a biased or misspecified model is preferred upon an actual better suited model.

\renewcommand{\arraystretch}{0.9}
\begin{table}[h]
\centering
	\captionabove{CRPS-H, CRPS-W, ES, VS and CRPS-U of the two benchmark and six copula-based models. The significantly best models according to the Diebold-Mariano test for every score are highlighted.}
	\label{t:scoreHourPV}
	\begin{tabular}{|	l	||	l	|	l	|	lll	|}
	\hline
		\textbf{model}&\textbf{CRPS-H}&\textbf{CRPS-W}&\textbf{ES}&\textbf{VS}&\textbf{CRPS-U}\\
	\hline
	\hline
HS&1&1&1&1&1\\
DA&0.8862&0.8532&0.5148&1.0895&0.8818\\
\hline
C1-Gumbel&\textbf{0.8601}&0.8807&\textbf{0.5017}&\textbf{0.8421}&\textbf{0.8749}\\
C1-Gaussian&\textbf{0.8594}&0.8789&\textbf{0.5025}&\textbf{0.8428}&0.8878\\
C1-BB1&\textbf{0.8581}&0.8647&\textbf{0.5023}&\textbf{0.8436}&0.8828\\
\hline
C2-Gumbel&\textbf{0.8577}&0.8532&\textbf{0.5019}&\textbf{0.8407}&\textbf{0.8739}\\
C2-Gaussian&\textbf{0.8596}&0.8466&\textbf{0.5034}&\textbf{0.8427}&0.884\\
C2-BB1&\textbf{0.8592}&\textbf{0.8386}&\textbf{0.502}&\textbf{0.8435}&0.884\\
	\hline
	\end{tabular}
\end{table}

Looking at the results we see that all copula-based models outperform the two benchmark models according to the CRPS-H with no significant differences between the various copula models. Thus we cannot distinguish the models due to the hourly marginals. A different picture emerges looking at the further scoring rules. The second benchmark model DA and the three C2-copula approaches that include daily dependencies by a Markov tree show the best scores for the scoring rule CRPS-W with the BB1 copula being significantly the best one. This is not surprising as this scoring rule emphasizes the dependencies and the BB1 copula is the only one that takes into account upper as well as lower tail dependence. These tail dependencies are important for a good estimation of the tails of the distribution of the weekly aggregated data. The general superiority of copula models compared to the benchmark models according to the dependence structure can be detected with the variogram score VS. The lack of an upper tail dependence of the Gaussian copula becomes apparent in the score CRPS-U. Here, the approaches with the Gumbel copula perform significantly better. 

\subsection{Evaluation results - upper and lower bounds}
\begin{figure}[h]
	\begin{center}
		\includegraphics[width=\textwidth]{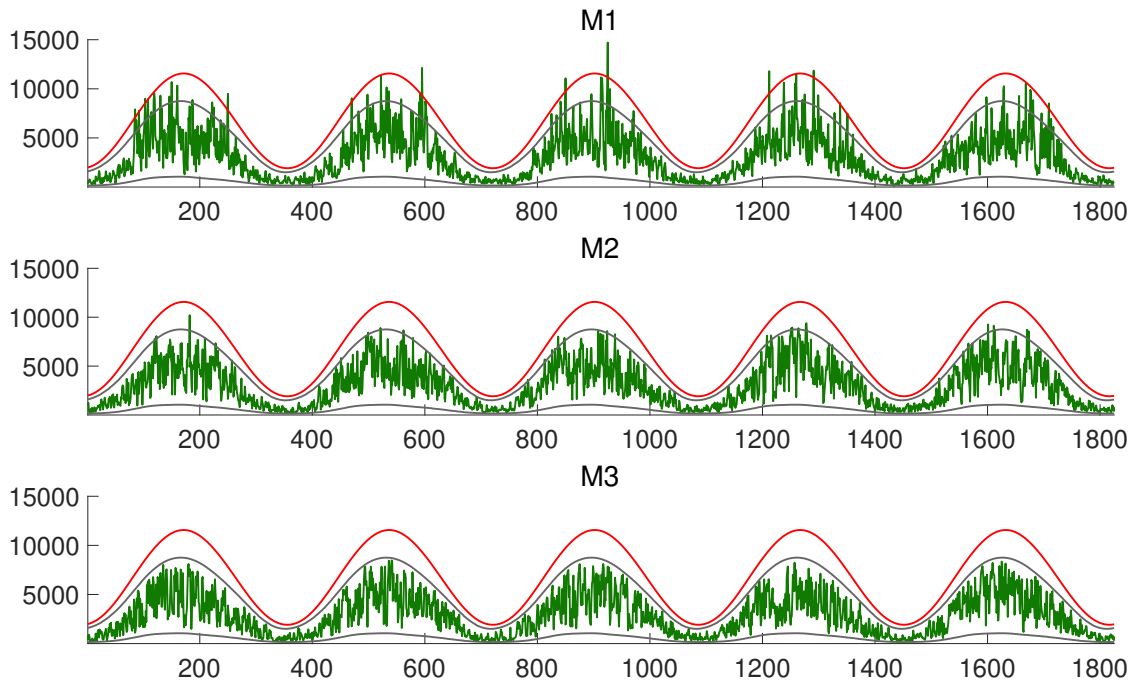}
	\end{center}
	\caption{Simulation paths with daily GHI for five years with upper and lower bounds (grey) and TOA (red) for all three model approaches.}
	\label{simulationsB}
\end{figure}
We also want to justify the need for an estimation method of the upper and lower bound. To do so, we introduce a simple stochastic model for daily GHI. Let $I_d$ be the daily GHI and consider the model equation
\begin{align*}
	h(I_d)=\Lambda_d+R_d
\end{align*}
with a link function $h$, a deterministic component $\Lambda_d$ and a stochastic component $R_t$. Such a model is used e.g. in \cite{benth2017} and \cite{veraart2015}. We estimate the deterministic component $\Lambda_d$ with a truncated Fourier series with orders $p=q=2$ and choose a classical ARMA(1,1)-model with skewed normal distributed innovations for the stochastic component.\\
We consider three different assumptions regarding the bounds of the irradiation. As a first case (M1) we assume no upper bound and therefore have $h_1:(0,\infty)\rightarrow\mathbb{R}$ and choose the logarithm as the link function like it is done in \cite{benth2017}. The natural upper bound TOA is chosen in the second approach (M2) $h_2:(0,\overline{g}^{toa}_d)\rightarrow\mathbb{R}$ and our estimated bounds in the third (M3) $h_3:(\overline{g}_d^{^-},\overline{g}_d^{^+})\rightarrow\mathbb{R}$. We denote the daily sums of the TOA and the estimated bounds by $\overline{g}^{toa}_d$ and $\overline{g}_d^{^\pm}$. We use the logit link for $h_2$ and $h_3$. In figure \ref{simulationsB} simulation paths for five years are shown together with our estimated upper and lower bounds and the TOA for all three assumptions regarding the bounds. A visual inspection already reveals the weaknesses of the first two choices of bounds. There are even exceedances of the natural upper bound TOA in the first approach and there are still exceedances of our estimated bound in the second approach. These exceedances lead to physically impossible values or to values that are considerably higher than historical minimum or maximum values. Thus, an interpretation and a reliable application of the model is not possible. This is also true if local or global PV yields are chosen instead of irradiations. For example a high infeed of PV energy can lead to negative electricity prices in the German market. Without a reliable upper bound of the PV infeed price models are prone to errors as they would produce points in time with too extreme negative prices. These shortcomings can also be underpinned by a look at a quantile weighted CRPS with a special emphasis of the tails (see \cite{gneiting2011}). The original formulation of the CRPS using the quantile decomposition is generalized by adding a nonnegative weight function $v$ on the unit interval giving different weights to different quantiles:
\begin{align*}
	\mathrm{CRPS^{QW}}(F,x)&=\int_0^1 2(\mathds{1}_{\{x<F^{-1}(\tau)\}}-\tau)(F^{-1}(\tau)-x)v(\tau)d\tau.
\end{align*}
Particular regions of the CRPS can be emphasized or down weighted with the weight function $v$. For $v\equiv 1$ we obtain the classical formulation of the CRPS. Here we use the weight functions
\begin{align*}
	v_1(\tau)&=(2\tau-1)^2,\\
	v_2(\tau)&=\mathds{1}_{\{\tau\leq 0.05\}}\quad \text{and}\\
	v_3(\tau)&=\mathds{1}_{\{\tau\geq 0.95\}}
\end{align*}
to put a special emphasis on both tails and the left and right tail of the distribution separately. Results of the three approaches are shown in table 4.4. The scores confirm that the use of our upper and lower bounds improves the model. 
\begin{table}[h]
\centering
	\captionabove{Quantile weighted CRPS for the three approaches, significantly best models according to a Diebold-Mariano test are highlighted}
	\begin{tabular}{|	l	||	lll		|}
	\hline
		&\textbf{CRPS-}$v_1$&\textbf{CRPS-}$v_2$&\textbf{CRPS-}$v_3$\\
	\hline
	\hline
		M1&1&\textbf{1}&1\\
		\hline
		M2&0.9663&1.0109&0.7317\\
		M3&\textbf{0.9513}&\textbf{0.9991}&\textbf{0.6288}\\
	\hline
	\end{tabular}
\end{table}

\section{Conclusion and Outlook}
In this paper we introduced probabilistic models for medium-term global horizontal irradiations using copula-based time series models with beta distributed marginals combined with an estimation procedure for the time varying upper and lower bounds. We have seen that the estimated lower and upper bounds of GHI outperform bounds that are frequently used in literature. Moreover, models using a Gumbel copula or a BB1 copula that allow for  tail dependence yield better probabilistic forecasts compared to models based on Gaussian copulas as evaluated with different scoring rules. We introduced new scoring rules based on the CRPS of functionals that have practical relevance in applications like the sizing of a solar accumulator or grid size planning.\\
The introduced model has many interesting applications that can be studied in the future. Combined with a load profile from an industrial company we would be able to analyze the effect of a planned PV plant on the load profile and its financial benefits or the fair value of a purchasing power agreement (PPA). Moreover, this model can be the basis for planning the optimal operation of a battery storage and thus the determination of the value of a battery storage and its effects on the grid load. Applying similar models to global PV yields of an area and using an additional model for electricity prices we could calculate future market value factors of PV under different assumptions on the development of installed capacities. These insights could help to evaluate the future value of PV energy projects and provides a great support for individual decision makers planning a PV plant as well as for political decision makers planning
laws for subsidies of renewable energies. 


\end{document}